\documentclass[12pt]{article}

\usepackage{enumerate}
\usepackage{amssymb}
\usepackage{amsmath}
\usepackage{amsfonts}
\usepackage{color}
\usepackage{verbatim}
\usepackage{graphicx}

\begin{document}

\begin{titlepage}

\begin{flushright}
\end{flushright}
\vskip 2.5cm

\begin{center}
{\Large \bf Quantum Effects of a Spacetime Varying $\alpha$ on the Propagation of Electrically Charged Fermions}
\end{center}

\vspace{1ex}

\begin{center}
{\large Alejandro Ferrero\footnote{{\tt alejandro.ferrero@gmail.com}; current
mailing address: Sabadell 1, Barcelona, Spain 08860}
and Brett Altschul\footnote{{\tt baltschu@physics.sc.edu}}}

\vspace{5mm}
{\sl Department of Physics and Astronomy} \\
{\sl University of South Carolina} \\
{\sl Columbia, SC 29208} \\
\end{center}

\vspace{2.5ex}

\medskip

\centerline {\bf Abstract}

\bigskip

A spacetime-varying fine structure constant $\alpha(x^\mu)$ could generate quantum
corrections in some of the coefficients of the
Lorentz-violating standard model extension (SME)
associated with electrically charged fermions. The quantum corrections depend on
$\partial_\mu\alpha$, the spacetime gradient of the fine structure constant.
Lorentz-violating operators involving fermions arise from the one-loop corrections
to the quantum electrodynamics (QED) vertex function and fermion self-energy. Both 
$g^{\lambda\mu\nu}$ and $c^{\mu\nu}$ terms are generated, at
${\cal O}(\partial_\mu\alpha)$ and
${\cal O}[(\partial_\mu\alpha)^2]$, respectively.
The $g^{\lambda\mu\nu}$ terms so generated are different in the vertex and
self-energy, which represents a radiatively induced violation of gauge invariance.

\bigskip

\end{titlepage}

\newpage

\section{Introduction}

Lorentz symmetry violations~\cite{ref-reviews} and spacetime-dependent
fundamental constants are two forms of fundamentally new physics beyond the standard
model. If compelling evidence for either of these phenomena were uncovered, it
would be a discovery of tremendous importance and could tell us a great deal about the structure of yet-to-be-discovered fundamental physics such as quantum gravity.

In recent years, there has been a significant amount on interest in both of these
possibilities. Interest in Lorentz violation, in particular, increased markedly after
the release of preliminary results from the OPERA experiment~\cite{ref-opera},
which appeared to show faster-than-light propagation of neutrinos. Although the
comparatively large Lorentz violation first reported by OPERA was eventually found
to have been an experimental artifact, the whole sequence of events has brought
new attention to the study of Lorentz violation in quantum theory.

While there is a long history of experimental tests of Lorentz symmetry, systematic
studies of the subject really only began about fifteen years ago. Since then, many
strong constraints have been placed on the leading order effects of the various
forms of Lorentz violation that could exist in effective quantum field theories and
other theoretical frameworks. Over approximately the same period, there has also been
significant interest in testing whether fundamental physical constants, such as the
fine structure constant $\alpha=\frac{e^{2}}{4\pi}$ or the electron-proton mass ratio
are truly
constants in time or whether they may have slight spacetime variations in their
values~\cite{ref-fischer,ref-gould,ref-tzanavaris}

If the fundamental couplings of the standard model are actually varying, this could
be closely related to the similarly exotic phenomenon of Lorentz violation.
In particular, if $\alpha$ is function that depends on spacetime
coordinates $\alpha(x^\mu)$, there is naturally a preferred spacetime direction,
$\partial_{\mu}\alpha$. If $\alpha$ varies purely in time, this preferred direction
violates boost invariance; in frames where $\alpha$ also depends on spatial
position, rotation invariance is also lost.

In this paper, we shall investigate the effects of a varying $\alpha$ on the
propagation and interaction of charged fermions. We shall adopt an essentially
minimal model, in which standard model Feynman rules are used to calculate the rates
for various radiative processes, with the sole modification that the coupling
constant $e$ is a function of the spacetime coordinates.
Similar calculations have already been performed in the photon sector with a
time-dependent $\alpha(t)$~\cite{ref-ferrero}; these demonstrated that
the interaction of photons with  virtual fermion-antifermion pairs 
could induce violations of Lorentz and gauge invariance.
However, these effects did not arise at $\mathcal{O}(\dot\alpha)$ but only at
$\mathcal{O}(\dot\alpha^2)$ and $\mathcal{O}(\ddot{\alpha})$, and so they would be
extremely strongly suppressed for physically allowed variations in $\alpha$.
In this paper, we shall neglect any effects that involve multiple derivatives
acting on $\alpha$; however, we shall include some $\mathcal{O}[(\partial_{\mu}
\alpha)^{2}]$ calculations, when they represent straightforward generalizations of
$\mathcal{O}(\partial_{\mu}\alpha)$ techniques.

Measurements of $\frac{\dot{\alpha}}{\alpha}$ have been done in different
ways: with pairs of precision spectroscopy experiments done years
apart~\cite{ref-fischer},
by analyzing the production rates for certain isotopes in natural 
reactors~\cite{ref-gould, ref-gould1}, and by observing spectra from cosmologically distant
sources like quasars~\cite{ref-tzanavaris}, among others.
The bounds that have resulted from these experiments are typically at the $\left|\frac{\dot{\alpha}}{\alpha}\right|<10^{-14}$
yr$^{-1}$ level for measurements of the present rate of change and a comparable
$\left|\frac{\Delta\alpha}{\alpha}\right|<10^{-5}$ level over cosmological time scales
set by the inverse of the Hubble constant $H$. Measurements
on spatial variations in $\alpha$ have also been studied using very distant 
sources~\cite{ref-webb1,ref-webb4}, where dipole and dipole plus monopole
variation patterns have been analyzed.

The cosmological searches for evidence of a nonzero $\partial_{\mu}\alpha$
benefit from very long photon propagation times. Radiation emitted in an earlier
epoch preserves information about the laws of the physics at the time of the
emission, and (apart from the cosmological Doppler shift) relatively little happens
to this radiation during the time it is propagating. Measurements involving the
propagation of charged
fermions are a much more difficult alternative, because the fermions interact very
easily with their environment. On the other hand, there are a number of efficient
ways to place constraints on Lorentz violation in the propagation of charged
fermion species; the most studied limits are on the Lorentz-violating parameters
denoted $c^{\mu\nu}$ in the electron sector~\cite{ref-stecker,ref-altschul7,
ref-hohensee1,ref-altschul20,ref-boquet,ref-altschul22}, which are directly related
to  the maximum speeds that electrons and positrons can achieve.

The $c^{\mu\nu}$ terms are the simplest Lorentz-violating terms that one can
introduce into an effective quantum field theory containing fermions. However,
the full effective field theory also contains many other kinds of operators. This
theory is known as the standard model extension (SME), and each Lorentz-violating
operator it contains is parameterized by a tensor-valued coefficient like
$c^{\mu\nu}$~\cite{ref-kost1,ref-kost2}.
These coefficients can be envisioned as (approximately constant)
background tensor fields, to which the standard model fields are coupled. In
addition to the spin-independent $c^{\mu\nu}$ coefficients in the fermion sector of the
SME, there are also a number of Lorentz violation coefficients that parameterize
spin-dependent forms of Lorentz violation. In this paper, we shall be particularly
interested in the spin-dependent $g^{\lambda\mu\nu}$ terms. Some of these can be
constrained (in combination with other terms) using laboratory experiments with
polarized electrons; however, some of the effects related to $g^{\lambda\mu\nu}$ are
harder to constrain, either because they would require observations of
astrophysical radiation sources with strongly polarized electron populations, or
because physically distinguishable effects only appear at second order in the
$g^{\lambda\mu\nu}$ coefficients.

In this paper, we shall present the results of our main calculations, which concern
radiation corrections to SME coefficients induced by the presence of a varying
$\alpha(x^{\mu})$, in sections~\ref{sec-results1} and~\ref{sec-results2}.
Section~\ref{sec-results1} discusses the corrections to operators that appear in the
pure fermion propagation Lagrangian, and section~\ref{sec-results2} addresses the
corrections to the fermion-photon interaction.
Our conclusions and outlook are presented in section~\ref{sec-concl}.

\section{Radiative Corrections to Fermion Propagation}

\label{sec-results1}

If $\alpha$ is variable on short spacetime scales, then radiative corrections, which
generally involve the production of virtual particles at a spacetime point $z_{1}$
and reabsorption or annihilation
at a different point $z_{2}$, will involve values of the
coupling $e$ at different spacetime points. This can generate Lorentz-violating
radiative corrections. The fermion self-energy and vertex corrections will include
SME-type operators that depend on $\partial_{\mu}\alpha$.

Which SME operators can be generated in this way is determined by the discrete
symmetry properties of the theory. The time derivative of a fundamental constant
(e.g. $\dot{\alpha}$) is odd under time reversal but even under parity and charge
conjugation; hence it is odd under the combined operation of CPT. However, many SME
operators violate CPT symmetry as well as Lorentz symmetry. (In fact, because of the
CPT theorem, the SME is actually the most general local, stable effective field
theory with standard model fields that allows CPT violation.)
The complementary phenomenon of spatially varying
constants is also odd under CPT, but odd under parity and even under time reversal and charge conjugation. Since the coefficients 
$g^{\lambda\mu\nu}$ violate both Lorentz and CPT symmetries, a spacetime-varying
$\alpha$ may
generate quantum corrections to some of these coefficients, and any new effects can
appear at $\mathcal O(\partial_\mu\alpha)$.
However, any corrections to the coefficients
$c^{\mu\nu}$ must depend on even powers of $\partial_\mu\alpha$, because $c^{\mu\nu}$
is Lorentz violating but CPT invariant. Hence, no corrections to $c^{\mu\nu}$ are 
expected at $\mathcal{O}(\partial_\mu\alpha)$.
By considering all the discrete symmetries of the other renormalizable SME operators
(which are discussed in~\cite{ref-kost4,ref-altschul8}), it is similarly possible
to rule out leading-order corrections to any operators except those described
by the $g^{\lambda\mu\nu}$.

We now turn to the evaluation of the radiative corrections to the $g^{\lambda\mu\nu}$
operators that control fermion propagation.
Following the same approach described in~\cite{ref-ferrero}, we will assume that
the theory will be defined by its Feynman rules. Because the coupling constant, and thus 
the Feynman rules, will be spacetime dependent, we shall set up the rules
in configuration space. The key element is the vertex, which is the only graph that
is modified under the introduction of a spacetime dependent $\alpha$. We could also use a field redefinition
of the form $A'^\mu=eA^\mu$ to eliminate the spacetime dependence in in the vertex; however, new terms in the
photon propagator appear and the calculations become more complicated. A fully
consistent theory might induce other changes to the Feynman rules,
but we want to consider only those effects which are
absolutely necessary consequences of having
a spacetime dependent $\alpha$. As usual, the fermion and photon 
propagators are given respectively by
\begin{eqnarray}\label{eq-1}
S_F(x-y)\!\!\!&=&\!\!\!\int\frac{d^4q}{(2\pi)^4}e^{-iq\cdot(x-y)}\frac{i(\not\!q+m)}{q^2-m^2+i\epsilon}\\
D_F^{\mu\nu}(x-y)\!\!\!&=&\!\!\!\int\frac{d^4q}{(2\pi)^4}\frac{-ie^{-iq\cdot(x-y)}}{q^2+i\epsilon}
\left(\eta^{\mu\nu}-\xi\frac{q^{\mu}q^{\nu}}{q^2}\right)
\end{eqnarray}
where $\xi$ is the standard gauge-dependent term. (We retain the gauge freedom of
$\xi$ even though the theory, with its time variable coupling $e$ does not have a
conserved charge, and so we cannot necessarily expect to find gauge invariant final
results.)
The electromagnetic vertex will be given by
\begin{eqnarray}\label{eq-2}
\includegraphics[scale=1.0]{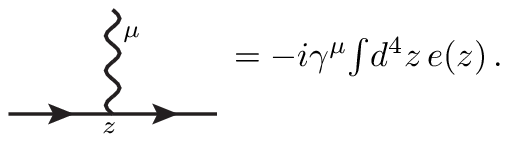}
\end{eqnarray}
The spacetime dependent $e(z)$ will be responsible for all the new effects.
We will assume that this dependence only comes from $\alpha$---thus keeping
$\hbar$ and $c$ constants. Under these assumptions, a spacetime dependent $e(z)=\sqrt{4\pi\alpha(z)}$ will be 
written, in terms of $\partial_\mu\alpha$
as
\begin{equation}\label{eq-2a}
\frac{e^2(z)}{\sqrt{4\pi\alpha_0}}=1+
\frac{1}{2}\frac{\partial_\mu\alpha}{\alpha_0}z^\mu
-\frac{1}{8}\left(\frac{\partial_\mu\alpha}{\alpha_0}z^\mu\right)^2
+\mathcal{O}\left[\partial_{\mu}\partial_{\nu}\alpha,(\partial_\mu\alpha)^3\right],
\end{equation}
where $\alpha_0$ represents the value of the fine structure constant at a spacetime
reference point.
In order to study the leading effects on the fermion propagation, we will start by studying
the fermion self-energy; it is represented by the Feynman diagram
\begin{center}
\includegraphics[scale=0.7]{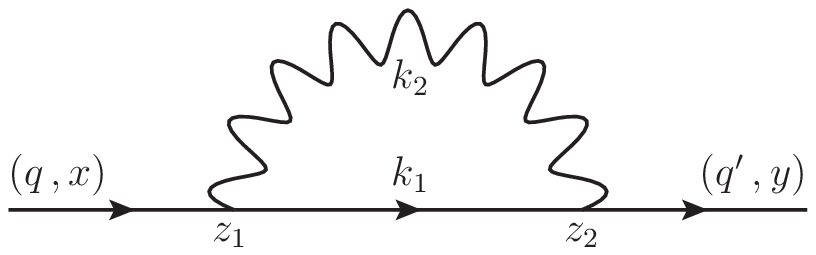}
\end{center}
where $x$ and $y$ represent the initial and final spacetime coordinates. 

Including the external legs, the formal expression for the one-loop correction to the fermion propagator is 
\begin{eqnarray}\label{eq-3}
\langle \psi(x)\bar\psi(y)\rangle\!\!\!&=&\!\!\!\int\frac{d^4q}{(2\pi)^4}
\frac{d^4q'}{(2\pi)^4}\frac{d^4k_1}{(2\pi)^4}\frac{d^4k_2}{(2\pi)^4}d^4z_1d^4z_2
\hat S(q)e^{-iq\cdot(x-z_1)}\nonumber\\
\!\!\!&&\!\!\!\times\bigg\{\big[-ie(z_1)\gamma_\mu\big]\frac{i(\not \!k_1+m)}{k_1^2-m^2}e^{-ik_1\cdot(z_1-z_2)}
\big[-ie(z_2)\gamma_\nu\big]\nonumber\\
\!\!\!&&\!\!\!\times\frac{-i}{k_2^2}\Big(\eta_{\mu\nu}-\xi\frac{k_2^\mu k_2^\nu}{k_2^2}\Big)e^{-ik_2\cdot(z_2-z_1)}\bigg\}
\hat S(q')e^{-iq'\cdot(z_2-y)},
\end{eqnarray}
where $\hat S(q)=\frac{i}{\not \,q-m}$. After a change in variables, $k=k_1$ and $p=k_1-k_2$, and integration over $k$, 
eq.~(\ref{eq-3}) can be brought into the form
\begin{eqnarray}\label{eq-4}
\langle \psi(y)\bar\psi(x)\rangle\!\!\!&=&\!\!\!\int\frac{d^4q}{(2\pi)^4}
\frac{d^4q'}{(2\pi)^4}\frac{d^4p}{(2\pi)^4}d^4z_1d^4z_2
e^{-iq\cdot(x-z_1)}e^{ip\cdot(z_2-z_1)}e^{-iq'\cdot(z_2-y)}\nonumber\\
\!\!\!&&\!\!\!\times\frac{e(z_1)e(z_2)}{e^2}\hat S(q)[-i\Sigma_2(p)]\hat S(q'),
\end{eqnarray}
where $-i\Sigma_2(p)$ represents the usual one-loop self-energy. 
The evaluation of eq.~(\ref{eq-4}) requires special attention because the external legs given by
$\hat S(q)$ and $\hat S(q')$, and the fermion self-energy do not necessarily commute.
Using the spacetime dependence of $e(z_1)$ and $e(z_2)$ introduced in eq.~(\ref{eq-2a}), 
and after carrying out the integrals over $z_1$, $z_2$, $p$ and $q'$ and further simplifications, we find
\begin{eqnarray}\label{eq-5}
\langle \psi(y)\bar\psi(x)\rangle\!\!\!&=&\!\!\!
\left[1+\frac{1}{2}\frac{\partial_\mu\alpha}{\alpha_0}(x+y)^\mu\right]\int\frac{d^4q}{(2\pi)^4}
e^{-iq\,\cdot(x-y)}\hat S(q)\left[-i\Sigma_2(q)\right]\hat S(q)\nonumber\\
\!\!\!&+&\!\!\!\frac{1}{2}\frac{\partial_\mu\alpha}{\alpha_0}
\int\!\frac{d^4q}{(2\pi)^4}e^{-iq\,\cdot(x-y)}
\hat S(q)\frac{q_\alpha\sigma^{\alpha \mu}}{q^2-m^2}\!\left[-i\widetilde\Sigma_2(q)\right]\!\hat S(q)\nonumber\\
\!\!\!&+&\!\!\!\frac{1}{8}\frac{\partial_\mu\alpha}{\alpha_0}\frac{\partial_\nu\alpha}{\alpha_0}
\int\frac{d^4q}{(2\pi)^4}e^{-iq\,\cdot(x-y)}
\hat S(q)\Big[\partial_{q_\mu}\partial_{q_\nu}\big\{\!-i\Sigma_2(q)\big\}\Big]\phantom{n}\!\!\!\hat S(q),
\end{eqnarray}
where (using a $d=4-\epsilon$ dimensional regularization scheme)
\begin{eqnarray}\label{eq-6}
\widetilde\Sigma_2(q)\!\!\!&\equiv&\!\!\!\frac{2me^2}{(4\pi)^2}\int_0^1dx
\frac{\Gamma(\epsilon/2)}{\Delta^{\epsilon/2}}\bigg\{
\Big(1-\frac{_1}{^2}x\xi\Big)\big[(4-\epsilon)-(2-\epsilon)x\big]+(4-\epsilon)x(1-x)\xi\bigg\}\nonumber\\
\!\!\!&-&\!\!\!\frac{2 me^2\xi}{(4\pi)^2}\int_0^1dx\frac{x(1-x)(1+x)q^2}{\Delta},
\end{eqnarray}
with $\Delta=(1-x)m^2-x(1-x)q^2$. 

Eq.~(\ref{eq-5}) can be divided into three parts. The first one 
is extracted from the first line and represents the usual correction for standard
quantum electrodynamics (QED); however, the fermion self-energy
is now spacetime dependent and modified by
\begin{eqnarray}\label{eq-6-a}
-i\Sigma_2(q)\rightarrow-\left\{\frac{e(x)e(y)}{e^2}+
\frac{1}{8}\left[\left(\frac{\partial_\mu\alpha}{\alpha_0}\right)(y-x)^\mu\right]^2
\right\}
i\Sigma_2(p).
\end{eqnarray}
The extra term arises because charge is no longer conserved; the expression simply
represents the self-energy, evaluated with the average value of $e$ in the interaction
region.

The term coming from the second line of eq.~(\ref{eq-5}) is of $\mathcal O(\partial_\mu\alpha)$
and introduces a quantum correction to the
SME coefficient $g^{\lambda\mu\nu}$ in the fermion kinetic sector.
The SME Lagrange density, including only the $c^{\mu\nu}$ and
$g^{\lambda\mu\nu}$ forms of Lorentz
violation is
\begin{eqnarray}
{\cal L} & = & \bar{\psi}(i\Gamma^{\nu}\partial_{\nu}-M)\psi \\
\Gamma^{\nu} & \supset & \gamma^{\nu}+c^{\mu\nu}\gamma_{\mu}+
\frac{1}{2}g^{\lambda\mu\nu}\sigma_{\lambda\mu}
\\
M & \supset & m.
\end{eqnarray}
The self-energy contains a (momentum-dependent) correction to the SME
$g^{\lambda\mu\nu}$,
\begin{eqnarray}\label{eq-7}
\frac{i}{2}\left(\Delta g^{\lambda\mu\nu}\right)\sigma_{\lambda\mu}q_\nu=
\frac{1}{2}\frac{\partial^\mu\alpha}{\alpha_0}
\frac{q^\alpha\sigma_{\alpha\mu}}{q^2-m^2}\big[-i\widetilde\Sigma_2(q)\big].
\end{eqnarray}
The largest contribution to eq.~(\ref{eq-6}) comes from its divergent part.
In the limit $q^2\ll m^2$, it induces a correction to the $g^{\lambda\mu\nu}$ term given by
\begin{equation}\label{eq-8}
\Delta g^{\lambda\mu\nu}=
\left\{\begin{array}{ll}
-\frac{3\alpha_0}{2\pi m}\left(\frac{\partial^\lambda\alpha}{\alpha_0}\right)
\eta^{\mu\nu}\ln\frac{\Lambda^2}{m^2}+\mathcal O\left(\frac{q^2}{m^2}\right)+\textrm{finite}, &\mu\neq\lambda\\
0,&\mu=\lambda
\end{array}\right.,
\end{equation}
where $\Lambda\sim e^{1/\epsilon}$
is an energy scale cutoff. This is the expected form for a logarithmic renormalization
of the $g^{\lambda\mu\nu}$ coefficient. Since the $g^{\lambda\mu\nu}$ have dimension
4 and are renormalizable in the standard way in the QED sector of the SME, this
new term can be straightforwardly incorporated into the renormalization of the
SME fermion sector~\cite{ref-kost4}.

However, terms with the structure of eq.~(\ref{eq-8}) can be eliminated from the purely fermionic part of the Lagrangian by a redefinition
of the fermion field, at least to first order in
$g^{\lambda\mu\nu}$~\cite{ref-collad}.
This implies that no observables related purely to fermion propagation
can have linear dependences on
the radiative induced contribution to $g^{\lambda\mu\nu}$. Nevertheless, the results coming from
the calculation of the electromagnetic vertex must also be analyzed in order to see if 
this field redefinition is still possible. 

The last term in eq.~(\ref{eq-5}) induces a correction to the $c^{\mu\nu}$
coefficients of the minimal SME. An analogous term was found in the photon
sector in~\cite{ref-ferrero}.
To see how this happens we must compute $-i\partial_{q_\mu}\partial_{q_\nu}\Sigma_2(q)$.
An interesting fact is that the divergent contributions are canceled when the derivatives over $q_\mu$ and $q_\nu$ are performed; hence, the largest
contribution is a finite term. Although we are more interested in the infinite contributions, it is interesting to see what kind of finite terms can arise in these calculations.
Using the formal expression of the photon self-energy and taking the limit $q^2\ll m^2$ we find 
\begin{eqnarray}\label{eq-9}
\partial_{q_\mu}\partial_{q_\nu}\Sigma_2(q)\!\!\!&=&\!\!\!\frac{\alpha_0}{2\pi m^2}\Bigg[\Big(2-\frac{_1}{^6}\xi\Big)m\eta^{\mu\nu}
-\Big(\frac{_2}{^3}-\frac{_{11}}{^{12}}\xi\Big)\big(\!\!\not\!q\eta^{\mu\nu}\!+q^\mu\gamma^\nu\!+q^\nu\gamma^\mu\big)\Bigg]
+\mathcal{O}\left(\frac{q^2}{m^2}\right)\nonumber\\
\end{eqnarray}
The terms proportional to $m$ and $\not\!\! q$ are small 
Lorentz-invariant renormalizations of
the fermion mass and field strength, respectively. 
The remaining contributions are Lorentz violating and induce quantum corrections to the $c^{\mu\nu}$ coefficients. Using
eqs.~(\ref{eq-5}) and (\ref{eq-9}), we find
\begin{eqnarray}\label{eq-10}
\Delta c^{\mu\nu}\!\!\!&=&\!\!\!\frac{\alpha_0}{12\pi m^2}
\left(\frac{\partial^\mu\alpha}{\alpha_0}\frac{\partial^\nu\alpha}{\alpha_0}\right)
\left(1-\frac{{11}}{{8}}\xi\right)+\mathcal O\left(\frac{q^2}{m^2}\right).
\end{eqnarray}

These results demonstrate that a spacetime varying $\alpha$ can generate
quantum corrections to minimal SME operators in the fermion sector, through virtual
fermion-photon loops in the fermion self-energy. At low energies, 
these effects are stronger for lighter fermions; because $\partial_{\mu}\alpha$ has
a positive mass dimension, it appears in conjunction with negative powers of the
fermion mass in radiative corrections to massless quantities such as
$c^{\mu\nu}$ and $g^{\lambda\mu\nu}$. This dependence also
has a natural physical interpretation. The radiative corrections are produced by
changes in the value of $\alpha$ between successive interaction vertices. The virtual
particles that exist between the interactions typically live a time $\sim\frac{1}{m}$,
so lighter species allow for more separation between the vertex locations and hence
larger effects.

The radiative corrections are also gauge
dependent; they depend on the gauge fixing parameter $\xi$ in an irreducible way.
This is actually unsurprising, since the minimal theoretical framework we have used,
which includes a varying coupling constant in the Feynman rules, is not gauge
invariant and does not include a conserved charge.
However, this may or may not be a full description of the physical behavior of
varying $\alpha$ theories. If there is real, physical variation in the fine
structure constant $\alpha$, there may or may not be a true conserved charge.
In more elaborate theories involving
varying couplings, in which additional charged fields with slowly varying expectation
values are introduced in order to rescue charge conservation, we anticipate that there
will be additional contributions to these radiative corrections, which will cancel
the gauge dependences found here.

\section{Radiative Corrections to the QED Vertex}

\label{sec-results2}

\begin{figure}
\centering
	\includegraphics[scale=0.6]{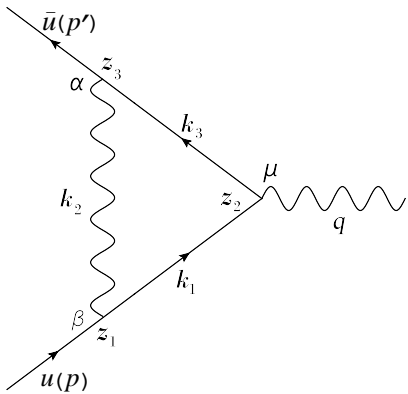}
	\caption{Leading quantum correction to the electromagnetic vertex}
	\label{fig-vert}
\end{figure}

We shall also find another form of gauge invariance
violation as we now turn to radiative corrections to the fermion-photon vertex
operator. The vertex correction
is the third key one-loop diagram in QED,
along with the fermion and photon self-energies, which have been looked at previously.
The structure of the vertex, which is shown in fig.~\ref{fig-vert}, that includes effects of a varying $\alpha$,
can be expressed as



\begin{eqnarray}\label{a3}
-ie\,\mathcal M^\mu\!\!\!&=&\!\!\!\int\frac{d^4k_1}{(2\pi)^4}
\frac{d^4k_2}{(2\pi)^4}\frac{d^4k_3}{(2\pi)^4}d^4z_1d^4z_2d^4z_3e^{-ip'\cdot z_3}\bar u(p')\Bigg\{[-ie(z_3)\gamma^\alpha]
\hat S(k_3)e^{-ik_3\cdot(z_2-z_3)}\nonumber\\
\!\!\!&&\!\!\![-ie(z_2)\gamma^\mu]\hat S(k_1)
e^{-ik_1\cdot(z_1-z_2)}[-ie(z_1)\gamma^\beta]\Bigg\}u(p)e^{ip\cdot z_1}
D_{\alpha\beta}(k_2)e^{iq\cdot z_2}e^{-ik_2\cdot(z_1-z_3)},\nonumber\\
\end{eqnarray}
which can be simplified to
\begin{eqnarray}\label{a3a}
-ie\,\mathcal M^\mu\!\!\!&=&\!\!\!-ie\int\frac{d^4k_1}{(2\pi)^4}
\frac{d^4k_2}{(2\pi)^4}\frac{d^4k_3}{(2\pi)^3}d^4z_1d^4z_2d^4z_3\frac{e(z_1)e(z_2)e(z_3)}{e^3}e^{-iz_1\cdot(k_1+k_2-p)}\nonumber\\
\!\!\!\!&&\!\!\!e^{-iz_2\cdot(k_3-k_1-q)}e^{-iz_3\cdot(p'-k_2-k_3)}
\bar u(p')v^\mu(k_1,k_2,k_3)u(p),
\end{eqnarray}
where
\begin{eqnarray}\label{eq-a34}
v^{\mu}(k_1,k_2,k_3)=-e^2D_{\alpha\beta}(k_2)\gamma^\alpha\hat S(k_3)\gamma^\mu\hat S(k_1)\gamma^\beta.
\end{eqnarray}
Eq.~(\ref{a3a}) can be solved up to $\mathcal{O}\left[(\partial_\mu\alpha)^2\right]$. However, the general result contains very long expressions, and the extraction of the 
finite contributions---as has been done for the fermion self-energy in the case
$q^2\ll m^2$---is extremely complicated. Thus, we shall only evaluate the infinite contributions, which are nonzero in this case.

In the SME, the coupling of Lorentz-violating fermions to the photon field occurs
through minimal coupling; the partial derivative $\partial_{\nu}$ is replaced by the
gauge covariant derivative $D_{\nu}=\partial_{\nu}+ieA_{\nu}$. This means that any
Lorentz violation coefficients that are expressed in $\Gamma^{\nu}$ actually appear
in both the free fermion sector and in the fermion-photon vertex. That the same
types of Lorentz violation appear in each place is required by gauge invariance, and
the Ward-Takahashi identity ensures that radiative corrections to the two operators
are related.
However, the minimal theory with a varying $\alpha$ does not enjoy gauge invariance, so
there can be entirely different radiatively induced values of $g^{\lambda\mu\nu}$
in the fermion propagation sector and in the electromagnetic interaction. We shall
demonstrate that the induced values of the $g^{\lambda\mu\nu}$ in the two
types of operators are indeed different; this result could have far-reaching
consequences, although the
physical consequences of this kind of gauge symmetry breaking have not been
investigated in detail.

Using the expressions
\begin{eqnarray}\label{eq-8b}
\partial_{p_\nu}\bar u(p)\!\!\!&=&\!\!\!\frac{1}{m}\bar u(p)\gamma^\nu,\,\,\,
\partial_{p_\nu}u(p)=\frac{1}{m}\gamma^\nu u(p),\,\,\,
\end{eqnarray}
which are easily obtained using the Weyl basis,
and the relation $\mathcal M^\mu=(2\pi)^4\delta(p'-p-q)\Delta\Gamma^\mu$, we find that the leading corrections to the electromagnetic vertex with a spacetime varying $\alpha$, as given in eq.~(\ref{a3a}), are
\begin{eqnarray}\label{eq-8na}
\Delta\Gamma^\mu=\frac{\alpha_0}{4\pi m}
\left(\frac{\partial_\nu\alpha}{\alpha_0}\right)\sigma^{\mu\nu}(1-\xi)\ln\frac{\Lambda^2}{m^2}+
\frac{\alpha_0}{24\pi m^2}
\left(\frac{\partial^\nu\alpha}{\alpha_0}\frac{\partial^\mu\alpha}{\alpha_0}\right)\gamma_\nu
(1-\xi)\ln\frac{\Lambda^2}{m^2}+\textrm{finite}.
\end{eqnarray}
If we denote the $g^{\lambda\mu\nu}$ that appears in the vertex as
$g_{\psi\psi A}^{\lambda\mu\nu}$, this includes a correction to the
$g_{\psi\psi A}^{\lambda\mu\nu}$ coefficients of
\begin{eqnarray}\label{eq-8n}
\Delta g_{\psi\psi A}^{\lambda\mu\nu}=
\left\{\begin{array}{rc}
-\frac{\alpha_0}{2\pi m}\left(\frac{\partial^\lambda\alpha}{\alpha_0}\right)
\eta^{\mu\nu}(1-\xi)\ln\frac{\Lambda^2}{m^2}+\textrm{finite}, & \lambda\neq\mu\\
0, & \lambda=\mu\end{array}\right.
\end{eqnarray}
Since there is a mismatch between the purely fermionic
$g^{\lambda\mu\nu}=g_{\psi\psi A}^{\lambda\mu\nu}$ and the
$g_{\psi\psi A}^{\lambda\mu\nu}$ that appears in the coupling, it is not possible to
eliminate a term with this Lorentz structure from the theory with simply a field
redefinition. There are expected to be real consequences for particle interaction
effects, although a more complete theory is needed to evaluate them, because a
specific gauge needs to be selected. (For the finely tuned value of $\xi=-2$---Yennie
gauge---the
infinite radiative corrections to the two $g^{\lambda\mu\nu}$ are actually equal;
however, that choice is not similarly satisfactory for other radiative correction
terms.)

There is also a correction to the vertex $c^{\mu\nu}$ given by
\begin{eqnarray}\label{eq-9qa}
\Delta c_{\psi\psi A}^{\mu\nu}\!\!\!&=&\!\!\!\!
\frac{\alpha_0}{24\pi m^2}
\left(\frac{\partial^\nu\alpha}{\alpha_0}\frac{\partial^\mu\alpha}{\alpha_0}\right)
(1-\xi)\ln\frac{\Lambda^2}{m^2}+\textrm{finite}.
\end{eqnarray}
This suggests the possibility of indirectly constraining a spacetime variation in $\alpha$, using experimental limits on the $c_{\psi\psi A}^{\mu\nu}$ appearing
in electron-photon interactions. 
The infinite term that arose from the calculation of the vertex diagram, describes the scale dependence of $c_{\psi\psi A}^{\mu\nu}$ under the renormalization group.
When $\xi<1$, the diagonal components $c^{00}$ and $c^{jj}$ are strictly positive. The signs of the non diagonal components $c^{0j}$ and $c^{ij}$, $i\neq j$ are positive when $\partial_0\alpha$ and $\partial_{i}\alpha$ are both positive or negative, and negative otherwise. The opposite behavior is expected when $\xi>1$.

\section{Outlook}

\label{sec-concl}

The most straightforward ways of looking for evidence of the Lorentz violation related
to a spacetime variable $\alpha$ involve studying fermion propagation.
The purely fermionic $c^{\mu\nu}=c_{\psi\psi}^{\mu\nu}$ modifies the dispersion
relations for electrons and other charged fermionic species.
Using the modified Dirac equation $(i\partial_\mu\Gamma^\mu-m)\psi$, with $\Gamma^{\nu}=\gamma^{\nu}+c^{\mu\nu}\gamma_{\mu}$ only,
the dispersion relation is given by $p^\mu p_\mu-m^2+2p_\mu p_\nu c^{\mu\nu}+\mathcal O\left[(c^{\mu\nu})^2\right]=0$.
If we define a coefficient $C$ according to $c^{\mu\nu}=C\frac{\partial^\mu\alpha}{\alpha_0}\frac{\partial^\nu\alpha}{\alpha_0}$, then
\begin{eqnarray}\label{eq-11}
p^\mu p_\mu-m^2+2C\frac{\partial^\mu\alpha}{\alpha_0}\frac{\partial^\nu\alpha}{\alpha_0}p_\mu p_\nu=0.
\end{eqnarray}
In a reference frame such that $p^\mu=(E,0,0,p)$, this reduces to
\begin{eqnarray}\label{eq-12}
E\approx\sqrt{p^2+m^2}\left\{1-C\left[\frac{\partial_0\alpha}{\alpha_0}-\frac{\partial_3\alpha}{\alpha_0}\frac{p}{\sqrt{p^2+m^2}}\right]^2\right\}.
\end{eqnarray}
Of course,
the full determination of the dispersion relation requires an exact calculation of
$C$, including the effects of finite terms.

Given the strong limits on $\partial_\mu\alpha$,
the quantum corrections coming from a varying $\alpha$ must be very small.
Terms of the $g^{\lambda\mu\nu}$ form should make the largest contributions to
observable effects, because the observable combination $g^{\lambda\mu\nu}_{\psi\psi}-
g^{\lambda\mu\nu}_{\psi\psi A}$ is nonzero at $\mathcal O(\partial_\mu\alpha)$.
In order to estimate the order of magnitude of the induced $\Delta g^{\lambda\mu\nu}$, we must assume an appropriate energy scale.
For $\Lambda\sim 10^{16}$ GeV, $\frac{3\alpha_0}{2\pi}\ln\frac{\Lambda^2}{m^2}\approx 0.3=\mathcal{O}(1)$. Then 
$\Delta g^{\lambda\mu\nu}\sim\eta^{\mu\nu}\left(\frac{\partial_\lambda\alpha}{\alpha_0}\right)t_C$, where $t_C$ is
the time interval associated with a fermion species' Compton wavelength; for electrons $t_{C}\sim 10^{-20}$ s. 
Using the existing limits on a time-variation of $\alpha$ of
$\left|\frac{\dot\alpha}{\alpha_0}\right|<10^{-14}$ yr$^{-1}$~\cite{ref-fischer,ref-gould,ref-tzanavaris,ref-gould1}, we find 
$\left|\Delta g^{0ij}\right|\lesssim 10^{-42}$.
Note that the effects derived from a spatial variation in $\alpha$ should lie near
the same order of magnitude.
Similarly, $\Delta c^{00}\sim
10^{-1}\left(\frac{\dot\alpha}{\alpha_0}\right)^2t_{C}^{2}$, so
$\Delta c^{00}\lesssim 10^{-85}$. These results imply that if any significant values for $g^{0ij}$ 
and $c^{00}$ are found, and more generally $g^{\lambda\mu\nu}$ and
$c^{\mu\nu}$, they should not be attributed to a time variation in $\alpha$ alone;
there would need to be other mechanisms operating to induce such effects.

In order to adapt these calculations to study a possible modification of the dispersion relation associated to neutrinos, we must
express the Feynman rules of the weak interactions in a framework where $\alpha$ is allowed to change over spacetime. In this case however, there are
other constants whose spacetime variation might also be analyzed. Using the appropriate set of variables, a similar procedure can be done. The mathematical structure of our results suggest that the relations $g^{\lambda\mu\nu}\propto m_f^{-1}$ and $c^{\mu\nu}_f\propto m^{-2}_f$ should also be found. 
To estimate the order of magnitude of this effect for neutrinos, we assume $m_{e}/m_{\nu_e}\sim 10^{10}$. Then, the induced effects on $g^{\lambda\mu\nu}$, and $c^{\mu\nu}$ on electron-type neutrinos could be enhanced by factors of $10^{10}$ and $10^{20}$ respectively (once the appropriate spacetime-varying variables are chosen). This is a large difference; however, values of $\vert g^{\lambda\mu\nu}_{\nu_e}\vert\sim 10^{-32}$ and $\vert c^{\mu\nu}_{\nu_e}\vert\sim 10^{-65}$ would still be far too small to have accounted for the results from the OPERA experiment.

In summary, we have studied the radiative corrections to the fermion sector caused by a
varying fine structure constant $\alpha$ in a simplified model. If $\partial_\mu{\alpha}\neq0$, the theory is not
invariant under Lorentz or CPT symmetry. Hence, there is
expected to be no symmetry preventing the appearance of Lorentz-
and CPT-violating terms in the effective action. If
only terms proportional to $\partial_\mu\alpha$ are considered, the one-loop fermion
self-energy does generate a quantum corrections to the SME coefficients $g^{\lambda\mu\nu}$, which are Lorentz, CPT, and gauge symmetry violating. At $\mathcal O\left[(\partial_\mu{\alpha})^2\right]$, the coefficients
$c^{\mu\nu}$ are also modified. However, all
these contributions are very small.

%

\end{document}